\documentclass[prb,twocolumn,floatfix,showpacs,superscriptaddress]{revtex4}
\usepackage{amsmath}
\usepackage{amsfonts}
\usepackage{graphicx}% Include figure files

\newcommand{\sigmaV}{{{\mbox{\boldmath$\sigma$}}}}

\font\bbfnt=msam10
\def\gsim{\,\hbox{\bbfnt\char'046}\,}

\def\ib{\mbox{\footnotesize \bf i}}
\def\jb{\mbox{\footnotesize \bf j}}
\def\kb{\mbox{\footnotesize \bf k}}
\def\qb{\mbox{\footnotesize \bf q}}
\def\Sb{\mbox{\boldmath $S$}}
\def\zero{\mbox{\boldmath $0$}}
\def\pib{\mbox{\boldmath $\pi$}}

\DeclareMathSymbol{\R}{\mathalpha}{AMSb}{'122}

\begin{document}

\title{Enhanced decoherence in the vicinity of a phase transition }
\author{S. Camalet}
\author{R. Chitra}
\affiliation{Laboratoire de Physique Th\'eorique de la Mati\`ere Condens\'ee, UMR 7600, Universit\'e Pierre et Marie Curie, Jussieu, Paris-75005, France}

\date{today}
\begin{abstract}
We study the decoherence of a spin-1/2 induced by an environment which is on the verge of a
continuous phase transition. We consider spin environments described  by the  
ferromagnetic and antiferromagnetic Heisenberg models on a square lattice. 
As is well known, these two dimensional systems undergo  a continuous phase transition 
at zero temperature, where, the spins order spontaneously.   
For weak coupling  of the central spin to these baths, we find that  
as  one approaches the transition temperature, critical 
fluctuations make the central spin decohere faster. 
 Furthermore, the decoherence is maximal at zero temperature as
 signalled by the divergence of the Markovian decoherence rate. 

\end{abstract}

\maketitle

The dynamics of a real system is determined not only by its internal Hamiltonian but also by 
its environment. Due to the coupling to the numerous environmental degrees of freedom, 
an initial pure quantum state of the system evolves into an incoherent mixture. 
This process, {\it decoherence}, is a major hurdle in the construction
of quantum computers  where sufficiently long coherence times of the qubits (the basic units of quantum 
information) are fundamental requisites.  
The recent spate of experimental work on  qubits has  generated a great deal of interest in the
question of decoherence induced by different environments. The most
studied environments are bosonic baths 
 modeled as   baths of harmonic 
oscillators \cite{Caldeira-Leggett}.  More recently, due to the numerous experimental realizations of qubits 
which involve real spin-1/2 objects \cite{EXPERIMENTS},  there has been an increasing focus on spin baths 
as a primary  source of decoherence.    

The rich and varied physics of spin systems make spin baths fundamentally interesting.
Spin baths comprising non-interacting spins have been among the first studied models to 
understand the decoherence process \cite{Zurek}. However, since the physical origin of the decoherence 
of a system  lies in the dynamical fluctuations of the bath degrees of
freedom to which it couples, we expect the resulting decoherence  to  reflect the
non-trivial nature of the fluctuations  
 induced by the interactions in the bath. 
From this perspective,  the vicinity of a continuous phase transition
is especially interesting given the existence of critical fluctuations.
Clearly, these divergent fluctuations are expected to have  dramatic consequences for the
 decoherence. Though some authors have studied spin bath models which exhibit
 continuous phase transitions, the effect of critical fluctuations has
been occulted either  because of the purely 
 mean field nature of the models \cite{italian-meanfields} or due to 
 the  temperature regimes considered \cite{Yuan,cb}.

 In this Letter, we address the issue of  the decoherence of a central
spin  weakly coupled to a spin bath which is on the verge of a phase
transition. We model the spin bath as a two-dimensional system of
Heisenberg spins with nearest neighbour interactions on a square
lattice. We consider both ferromagnetic and antiferromagnetic
interactions. In both cases, the spin bath undergoes spontaneous symmetry breaking at $T_c=0$ to a magnetic phase. 
These spin bath models are interesting for our purpose as their critical fluctuations are well understood \cite{Manousakis}. 
Studies on fluctuation free mean field models
 indicate that the decoherence time scale increases as temperature is
 lowered and the system undergoes a transition \cite{italian-meanfields}. Moreover, this result is in
 accord with the conventional wisdom that decoherence is minimal at $T=0$ and increases
 with increasing thermal fluctuations. 
 In what follows, we will show that  contrary to the above scenario, as one approaches the critical point
 at  zero temperature,  critical fluctuations result in a faster
decoherence of the central spin  with 
the decoherence being maximal at 
 zero temperature.

We consider a central  spin-1/2
coupled  isotropically  via a hyperfine like contact interaction to a 
spin-$S$ bath with uniform nearest neighbor interactions.  For simplicity,  we assume
that the central spin  has no internal dynamics.  The total Hamiltonian
describing the central spin and its environment is given by    
\begin{eqnarray}
H&=&H_I+H_B \nonumber \\
&\equiv&{\sigmaV} \cdot\sum_{\ib} \lambda_{\ib}  \Sb_{\ib}  
-J \sum_{( \ib  \jb )} \Sb_{\ib} \cdot \Sb_{\jb}
 \label{H}
\end{eqnarray}
where $\sigmaV$ are the Pauli matrices, $\Sb_{\ib}$ the spin operators of the bath spin 
at site $\ib$ and $( \ldots )$ denotes summation over nearest neighbor 
spin pairs of a square lattice with periodic boundary conditions. For the case of an electronic spin interacting
with nuclear spins through the  hyperfine contact interaction,
the coupling constant $\lambda_{\ib}$ is simply related 
to the electron envelope wave function at the site $\ib$. 
Antiferromagnetic interactions ($J<0$) and ferromagnetic interactions
($J>0$) lead to Neel order and ferromagnetic order respectively, at $T=0$.  
To study the decoherence for weak coupling to 
the spin bath and a finite temperature $T$, we use the 
resolvent operator approach \cite{ROM1,ROM2,nous} in conjunction with the
Schwinger boson technique. 
To simplify the calculation, we  assume a factorizable initial density matrix for the composite system, 
$\Omega=\rho(0)  \otimes \rho_B$ where $\rho(0)$ and
$\rho_B \propto \exp(-H_B/T)$  denote,  respectively, the  central spin state and the thermal equilibrium of the bath. 
We use units $k_B=\hbar=1$ throughout this paper.

Since the total Hamiltonian \eqref{H} is rotationally invariant and the finite temperature phase of the bath is isotropic, 
the central spin Bloch vector at time $t$ is related to its initial value as given by  
\begin{equation}
\langle \sigmaV \rangle(t)=r(t)\langle \sigmaV \rangle(0)  \label{defr}
\end{equation}
where $r$ is a scalar function of the time $t$ and $ \langle\ldots\rangle$ denotes
the average over the density matrix of the composite system.  
Due to the coupling to the bath, the function $r$ vanishes in the long time limit. 
To determine $r(t)$, it is useful to write the  reduced density matrix
$\rho(t)$ of the central spin as
\begin{equation}
\rho(t) = \frac{i}{2\pi} \int_{\R+i\eta} dz  e^{-izt}  
\mathrm{Tr}_B \left[  \left( z-{\cal L} \right)^{-1} \rho(0) \rho_B \right] \label{rhot} 
\end{equation}  
where $\eta$ is a real positive number, $\mathrm{Tr}_B$ denotes the partial trace over the bath degrees 
of freedom and ${\cal L}$ is the Liouvillian corresponding to the total  Hamiltonian $H$, i.e., 
${\cal L}A=[H,A]$ for any operator $A$.
Using the decomposition of the density matrix $\rho(t)$ in  the  basis of  $2\times 2$ matrices $\{ I, \sigma_\alpha \}$ 
and the projection operator technique explained in
Refs.~\onlinecite{ROM2} and \onlinecite{nous}, we obtain  from \eqref{rhot} 
the expression \eqref{defr} where 
\begin{equation}
r(t)= \frac{i}{2\pi} \int_{\R+i\eta} dz  \frac{e^{-izt}}{ z-\Sigma(z) } \label{M} .
\end{equation}  
\noindent
The self-energy $\Sigma$ is given by 
\begin{equation} 
\Sigma(z)= {\textstyle \frac{1}{2}} 
\mathrm{Tr} \left[ \sigma_\alpha \mathrm{Tr}_B 
\left[ {\cal L}_I (z-{\cal Q}\, {\cal L} \,{\cal Q})^{-1}{\cal L}_I \rho_B  \sigma_\alpha \right]\right] \label{Sigma}
\end{equation}
where $\sigma_\alpha$ is any Pauli matrix and ${\cal L}_I$ is the Liouvillian corresponding to the interaction 
Hamiltonian $H_I$. The projection operator ${\cal Q}$ is defined by its action on any operator $A$ as
${\cal Q} A  =  A - \mathrm{Tr}_B (A) \rho_B$. 
We note that for any arbitrary Hamiltonian, the decoherence of the central spin cannot be described 
by a single self-energy, the full time evolution of the state $\rho(t)$ is given by a $4 \times 4$ 
matrix function of the complex variable $z$. 
From this $4 \times 4$ matrix, it can be shown that the asymptotic behavior of  $\rho(t)$ is in general characterised 
by two different times, a decoherence time and a relaxation time which determines the time for thermal equilibration of 
the central spin \cite{QDS}. For the Hamiltonian \eqref{H}, the time-scales 
for relaxation and decoherence are  the same. However, for the generic
case of the central spin having its own internal dynamics, these two
times are in principle different. This generic case is studied in the
last part of the paper.

The decoherence in the weak coupling regime is determined by the lowest order contribution of the interaction 
Hamiltonian  $H_I$ to the self-energy $\Sigma$. The first term of this expansion is given by \eqref{Sigma} with  
${\cal L}$ replaced by the bath Liouvillian ${\cal L}_B$. 
Using the properties of the Pauli matrices, 
this second-order contribution to the self-energy  can be  written as 
\begin{equation}
\Sigma_2(z) = -i \frac{8}{3}  \sum_{\ib,\jb}\lambda_{\ib}\lambda_{\jb} \int_0^\infty dt\, e^{izt} \mathrm{Re} 
\langle \Sb_{\ib} (t) \cdot \Sb_{\jb} \rangle_B \label{Sigma2} 
\end{equation}       
where $\Sb_{\ib} (t)=\exp(iH_B t) \Sb_{\ib}\exp(-iH_B t) $ and $\langle A \rangle_B=\mathrm{Tr} (\rho_B A)$ 
denotes the thermal average  of any bath operator $A$. Neglecting higher order contributions to $\Sigma$ 
in \eqref{M} is equivalent to the Born approximation \cite{QDS}.  
It can be shown \cite{nous} that in  the weak coupling limit,  the expression \eqref{M} can be simplified to obtain
\begin{equation}
\ln r(t) \simeq - \frac{2}{\pi} \int dE\,  \frac{\sin (tE/2)^2}{E^2} \Gamma_2(E) \label{rapprox}
\end{equation} 
where $\Gamma_2(E)=-\mathrm{Im}  \Sigma_2(E+i0^+)$. 
Eq.\eqref{rapprox} describes the decoherence at all time scales. It yields the usual quadratic decrease $r(t) \simeq 1 
-  t^2 \int dE\,   \Gamma_2(E)/2\pi$ for very short times and Markovian decay $r(t)\simeq\exp[-\Gamma_2(0)t]$ for asymptotic times. 
The full time evolution of the central spin is determined by the
dynamic spin correlation function of the bath 
via the relations 
\eqref{rapprox} and \eqref{Sigma2}. Typically, dynamical correlations are rather difficult to calculate for spin systems. 
In the following, we  use the successful Schwinger boson mean field theory to evaluate the dynamical spin correlation of 
the bath described by $H_B$.
We will show that  the critical fluctuations in the bath  lead
to a  divergence of $\Gamma_2(0)$ at the transition, implying a faster than
exponential asymptotic decay of $r(t)$.
This divergence is merely a reflection of the divergence of the
 underlying correlation time in
the bath at the phase transition.

In the Schwinger representation \cite{AA1}, at every site ${\ib}$, the spin operators $\Sb_{\ib}$ are replaced by two bosonic operators
$a_{\ib}$ and $b_{\ib}$ with the constraint $a_{\ib}^\dag a^{\phantom{\dag}}_{\ib} + b_{\ib}^\dag b^{\phantom{\dag}}_{\ib}=2S $. 
In the ferromagnetic case, the correspondence relations read 
$
2S_{\ib}^z=a_{\ib}^\dag a^{\phantom{\dag}}_{\ib} - b_{\ib}^\dag b^{\phantom{\dag}}_{\ib} $ and 
$S_{\ib}^+=a_{\ib}^\dag b^{\phantom{\dag}}_{\ib}$.
In the antiferromagnetic case, due to the possibility of Neel ordering of spins,  two sub-lattices must be distinguished. 
On one of the sub-lattices, the spin and boson operators are related as above, while on the other  sub-lattice 
the correspondence relations take the form
$2S_{\ib}^z=b_{\ib}^\dag b^{\phantom{\dag}}_{\ib}-a_{\ib}^\dag a^{\phantom{\dag}}_{\ib}$ and 
$S_{\ib}^+=-b_{\ib}^\dag a^{\phantom{\dag}}_{\ib}$.   In both ferromagnetic  and antiferromagnetic 
cases, the Hamiltonian \eqref{H} can be interpreted as describing 
a spin-1/2  coupled to a  boson bath.   The resulting problem is  
nonetheless very different from the   standard spin-boson model for the following reasons: 
i)  the central spin couples to two species of bosons ii) the bath bosons interact with each other and  are subject 
 to constraints which conserve the  number of bosons. In the Schwinger boson mean field theory, 
 the local constraints on the bosons are replaced by a global constraint via a uniform chemical potential and the boson 
 Hamiltonian derived  from $H_B$ is studied using a Hartree-Fock mean field scheme  \cite{AA1,T,HF}. 
 We now present analytical results for the decoherence obtained within this theory in the low temperature 
 regime $T \ll |J|S$.

For the ferromagnetic Heisenberg model
 ($J>0$), the mean-field approximation for $H_B$  yields, up to a
constant energy,
$H_{MF}^{fm}=\sum_{\kb} \omega_{\kb} (a_{\kb}^\dag a^{\phantom{\dag}}_{\kb} 
+ b_{\kb}^\dag b^{\phantom{\dag}}_{\kb})$ where $a_{\kb}$ and $b_{\kb}$ are 
the Fourier transforms of $a_{\ib}$ and $b_{\ib}$, respectively. 
The  magnon dispersion is given by
\begin{equation}
\omega_{\kb}=2JQ(2-\cos k_x -\cos k_y) - \mu
\end{equation}
where $k_x$ and $k_y$ are the components of the wave-vector $\kb$. 
The mean field parameter $Q$  and the chemical potential $\mu$  are determined by  the self consistent equations 
\begin{eqnarray}
S &=& \frac{1}{N}\sum_{\kb} n_{\kb} \nonumber \\ Q &=& \frac{1}{N}\sum_{\kb} n_{\kb} \cos k_x \label{sceq}
\end{eqnarray}
where $n_{\kb}=[\exp(\omega_{\kb}/T)-1]^{-1}$ is the Bose occupation factor. Note that here the chemical 
potential $\mu$ is necessarily negative. At low temperatures, since  the above sums are dominated by the modes 
$\omega_{\kb} \ll T$,  $Q\simeq S$ and $\mu \to 0$ as $T\to 0$ to ensure  that 
the total particle number remains fixed.   
A more precise analysis of \eqref{sceq} gives $\mu \simeq -T \exp(-4\pi J S^2/T)$. The transition to the ordered state is explained 
by a  pseudo Bose condensation  of the bosons at the critical temperature \cite{HF}.      

Using the above results, we find the decoherence in the weak coupling regime is given by \eqref{rapprox}  with  
\begin{eqnarray}
\Gamma_2(E) &=& \frac{4\pi}{3N^2} \sum_{\kb,\qb}|\Lambda ({\kb}-{\qb})|^2  \delta (E+\omega_{\kb}-\omega_{\qb})
\nonumber \\  &&\times	(n_{\kb}+n_{\qb}+2n_{\kb}n_{\qb})  \label{Gammaferro}
\end{eqnarray}
where  $\Lambda ({\kb})=\sum_{\ib} \lambda_{\ib} \exp(i{\kb}\cdot{\ib})$. In accordance with
Ref.\onlinecite{AA1}, a multiplicative factor $2/3$ has been taken into account  so as to ensure 
 that the spin correlation function satisfies the correct sum rules.
In the zero temperature limit, since $n_k$ vanishes for $\omega_k \gg T$, the last term of \eqref{Gammaferro} is a Dirac function 
at $E=0$ whereas the first two terms converge to a continuous function of $E$ with the characteristic energy $JS$ 
and a finite value, $\Lambda (\zero)^2/3J$, at $E=0$. For times $t \ll (JS)^{-1}$, the sine function in \eqref{rapprox} can be expanded 
leading to the usual short-time quadratic decoherence. For  $(JS)^{-1} \ll t \ll T^{-1}$, the last term of \eqref{Gammaferro} 
is essentially a Dirac function in \eqref{rapprox} as it practically vanishes for $|E| \gsim  T$. Moreover, the contribution of the first two 
terms of \eqref{Gammaferro} to $\ln r(t)$, $-t \Lambda (\zero)^2 /3J$,  is negligible and then  
\begin{equation}
\ln r(t) \simeq -\frac{4}{3} (\Lambda_{fm} S t)^2  \label{Gauss}
\end{equation}
where $\Lambda_{fm}\equiv\Lambda (\zero)=\sum_{\ib} \lambda_{\ib}$. 
To obtain the decoherence for longer times, we evaluate $\Gamma_2(E)$ for energies $|E| \ll T$ using 
the approximations $\omega_{\kb}\simeq J S(k_x^2+k_y^2)$ and $\Lambda ({\kb}-{\qb}) \simeq \Lambda(\zero)$. 
We find
 \begin{equation}
\Gamma_2(E)=\frac{\Lambda_{fm}^2 T^2}{6\pi J^2 S^2| \mu|} \left|\frac{\mu}{E}\right|
\ln \left( 1+\left|\frac{E}{\mu}\right|\right). \label{Gammaapprox}
\end{equation}
\noindent
For times $T^{-1} \ll t \ll |\mu|^{-1}$, \eqref{Gammaapprox} results in a small positive correction to 
the Gaussian decay \eqref{Gauss},
 $
\ln r(t) +4 (\Lambda_{fm} S t)^2/3  \simeq 2 T (\Lambda_{fm} t)^2 \ln (T t)/3\pi J. 
$  
For longer times, $t \gg |\mu|^{-1}$, the decoherence is  Markovian with the rate $\Gamma_2(0)$.  
The non-Markovian corrections arising 
from the low energy behaviour of \eqref{Gammaapprox} are logarithmic, 
$\ln r(t) +  \Gamma_2(0) t \simeq (\Lambda_{fm} T/\pi J S\mu)^2 \ln(|\mu|t)/6$. 
As anticipated, the rate $\Gamma_2(0)$ diverges in the zero temperature limit, implying that the
asymptotic decoherence is Gaussian at $T=0$. Note that the cross-over from the Gaussian decoherence 
given by \eqref{Gauss} to the Markovian decoherence  is extremely long at low temperatures.

For the  antiferromagnetic bath, following the steps outlined in
Refs.~\onlinecite{AA1} and \onlinecite{HF},
 we obtain the following mean field Hamiltonian for the bath (up to a
constant)
$
H_{MF}^{afm}=\sum_{\kb} \omega_{\kb} (\alpha_{\kb}^\dag\alpha^{\phantom{\dag}}_{\kb} 
+ \beta_{\kb}^\dag\beta^{\phantom{\dag}}_{\kb})
$ 
where $\alpha_{\kb}$ and $\beta_{\kb}$ are  linear combinations of the  Fourier transforms of the original operators 
$a_{\ib}$ and $b_{\ib}$, respectively. The magnon dispersion is now given by
\begin{equation}
\omega_{\kb}= \sqrt{\mu^2 -A_{\kb}^2}
\end{equation}
where $A_{\kb}\equiv 2JQ(\cos k_x +\cos k_y)$ and the corresponding
 self consistency conditions are 
\begin{eqnarray}
S + \frac12 &=& \frac{1}{N} \sum_{\kb} \frac{\vert\mu\vert}{\omega_{\kb}}\left(n_{\kb}+\frac12\right)
\nonumber \\ 4JQ^2 &= & \frac{1}{N} \sum_{\kb} \frac {A_{\kb}^2}{\omega_{\kb}} \left(n_{\kb}+ \frac12\right) . \label{afmSelfcon}
\end{eqnarray}
  In the paramagnetic phase, $ T >0$,   there  exists  a  gap   $\Delta=(\mu^2 -16J^2
Q^2)^{1/2}$ in the magnon dispersion for 
$\kb = {\bf 0}$ and the Neel  ordering wavevector $\kb =
(\pi, \pi)$. An analysis of \eqref{afmSelfcon} shows that as $T\to 0$,
 $Q \to Q_0 \simeq S+0.08$  and the gap vanishes as 
$\Delta \simeq T\exp(-2 \pi |J| {\bar \rho}_s /T )$ where $|J|\bar{\rho}_s$ is the spin stiffness of the bath \cite{Manousakis}. 
The dimensionless parameter $\bar{\rho}_s$ depends only on $S$, $\bar{\rho}_s\simeq 0.176$ for $S=1/2$ and 
$\bar{\rho_s} \simeq S^2$ for $S \gg 1$.

A direct calculation of the dynamic spin correlation yields \cite{AA2}
\begin{widetext}
\begin{equation}
\Gamma_2(E) = \frac{\pi}{3N^2} \sum_{\kb,\qb}|\Lambda ({\kb}-{\qb})|^2 
  \sum_{\epsilon=\pm 1} \frac{3+\epsilon}{2} 
  \left[ \frac{\mu^2-A_{\kb}A_{\qb}}{\omega_{\kb}\omega_{\qb}}+\epsilon \right]
\left[ 2n_{\kb}n_{\qb}+n_{\kb}+n_{\qb} +  \frac{1-\epsilon}{2} \right]
\delta (|E|+\epsilon\omega_{\kb}-\omega_{\qb})
\label{Gammaantiferro}
\end{equation}
\end{widetext}
\noindent
 As in the ferromagnetic case, in the zero temperature limit, the sum of the terms $\propto n_{\kb} n_{\qb}$ 
 is a Dirac function at $E=0$ whereas the other terms converge to a continuous function of $E$ 
 with finite characteristic energy, $|J|Q_0$, and value at $E=0$, $2\Lambda (\pi,\pi)^2 {\bar {\rho_s}} /3 |J| Q_0^2$. 
Here also, for temperatures $T \ll |J|S$, this continuous function contributes to the decoherence 
only in the very short time regime $t \ll (|J|S)^{-1}$. To evaluate the decoherence for longer times,        
we remark that, for $T \ll |J|S$ and $|E| \ll |J|S$, the  sums over $\kb$ and $\qb$ in \eqref{Gammaantiferro} are
dominated by the vicinities of  $({\kb},{\qb})=(\zero,\pib)$ and $({\kb},{\qb})=(\pib,\zero)$ where $\pib=(\pi,\pi)$ 
permitting us to expand $\omega_{\kb}$ and $\omega_{\qb}$ to quadratic order in ${\kb}$ and ${\pib}-{\qb}$ 
or vice-versa in conjunction with the consistent approximations $|\Lambda ({\kb}-{\qb})|\simeq \Lambda ({\pib})$ and 
$\mu^2-A_{\kb}A_{\qb} \simeq 8|J|Q_0$. Using these, we find the decoherence is Markovian for times  $t \gg \Delta^{-1}$ 
with the rate 
\begin{equation}
\Gamma_2(0)=\frac{\Lambda^2_{afm} T^2}{3\pi J^2 Q_0^2\Delta}
\end{equation}
 where $\Lambda_{afm}\equiv\Lambda (\pib)$. For times $(|J|S)^{-1} \ll t \ll T^{-1}$,  taking into account 
 the terms with $\epsilon=-1$ in \eqref{Gammaantiferro} we obtain
\begin{equation}
\ln r(t) \simeq -\left( \Lambda_{afm} \frac{\bar{\rho_s}}{Q_0} t \right)^2  .
\end{equation}
As in the ferromagnetic case, the rate $\Gamma_2(0)$ diverges in the zero temperature limit and the asymptotic decoherence 
is Gaussian at $T=0$. We note that the
decoherence is faster as $S \to \infty$ i.e., when the spins become {\it
classical}.  In this case, since the associated scales $\mu, \Delta \to 0$, the
Markovian regime is not accessible and the decoherence is Gaussian
with a characteristic time  $ \propto S^{-1}$.

We observe that the decoherence is qualitatively similar 
for ferromagnetic and antiferromagnetic interactions in the bath but
with the important difference that 
$\ln r(t)$ is proportional 
to $\Lambda_{fm}^2=\Lambda (\zero)^2$ or to $\Lambda_{afm}^2=\Lambda (\pib)^2$, respectively. This can be understood as follows : 
an enhanced decoherence is observed in the vicinity of the transition
temperature only if the central 
spin couples to the critical mode of the bath. 
We now show  that this critical enhancement of the  decoherence is not
contingent on the absence of internal dynamics for the central spin.
 Due to the rotational invariance of  $H$,  any  intrinsic dynamic for
the central spin can be described by  the total Hamiltonian $H'=H+ \epsilon \sigma_z/2$. In this case, as mentioned earlier, 
the Markovian asymptotic behavior of the central spin state 
is characterized by two times \cite{QDS,Loss}, a relaxation time $T_1=\Gamma_2(\epsilon)^{-1}$ and a decoherence time $T_2$ given 
by $T_2^{-1}=T_1^{-1}/2+\Gamma_2(0)/2$. At the critical point,    for any value of $\epsilon$, $T_2$
vanishes and the resulting  decoherence 
is faster than exponential. On the other hand, the behavior of  $T_1$ 
depends on the value of $\epsilon$. 
For $|\epsilon| \ll |J|S$, the above study of the function $\Gamma_2$ shows that $T_1^{-1}$ reaches a maximum 
at $T \sim |\epsilon|$. This maximum grows as $\epsilon\to 0$ and we recover the 
critical enhancement of the decoherence for $\epsilon =0$, i.e., $ T_1=T_2 \to
0$. We finally remark that, in the low temperature limit, since  $Q$ is
essentially constant, the spin environments we consider are
practically  similar to baths of independent  but {\it conserved} bosons.

In conclusion, the two models studied in this paper clearly illustrate
the phenomenal impact of  fluctuations 
near a critical point 
on the decoherence and that to minimize decoherence it is essential to avoid continuous phase transitions in
the bath. We expect our conclusions to be generically valid for any
bath manifesting a continuous phase transition  provided the central spin couples to the relevant
critical modes. 
 A natural extension of our work  would be to study the decoherence and relaxation
induced by a higher-dimensional bath  in  the
ordered phase below $T_c \neq 0$ so as  to compare the weakly fluctuating limit $T \to 0$ and the strongly 
fluctuating limit $T \to T_c^{-}$. Given the richness of the decoherence studied in this paper, it would
be interesting to explore the decoherence induced by  baths  in  quantum
critical regimes.

\end{document}